\documentclass[conference]{IEEEtran}

\usepackage{multirow}

\usepackage{amsmath,amsfonts, amsthm, amssymb, bm, xcolor}
\usepackage{algorithm}
\usepackage{comment, tagging}
\def\BibTeX{{\rm B\kern-.05em{\sc i\kern-.025em b}\kern-.08em
    T\kern-.1667em\lower.7ex\hbox{E}\kern-.125emX}}

\usepackage[compress]{cite}
\usepackage[noend]{algpseudocode}
\usepackage{xspace, paralist, xcolor}
\PassOptionsToPackage{hyphens}{url}
\usepackage{float, graphicx, pstricks, pst-node,pst-plot}
\usepackage{subfloat, tabu, color, tabto}
\usepackage{subfiles, subcaption}

\usepackage{colortbl}
\usepackage{stackengine} 

\usepackage{array}
\newcolumntype{P}[1]{>{\centering\arraybackslash}p{#1}}
\newcolumntype{M}[1]{>{\centering\arraybackslash}m{#1}}
\usepackage{setspace}

\algnewcommand{\AND}{\textbf{and}\xspace}
\algnewcommand{\OR}{\textbf{or}\xspace}
\algnewcommand\algorithmicforeach{\textbf{for each}}
\algdef{S}[FOR]{ForEach}[1]{\algorithmicforeach\ #1\ \algorithmicdo}

\algrenewcommand\algorithmicindent{1.0em}%

\makeatletter
\algnewcommand{\LineComment}[1]{\Statex \hskip\ALG@thistlm \(\triangleright\) #1}
\makeatother

\newbox\statebox
\newcommand{\myState}[1]{%
    \setbox\statebox=\vbox{#1}%
    \edef\thealgruleheight{\dimexpr \the\ht\statebox+1pt\relax}%
    \edef\thealgruledepth{\dimexpr \the\dp\statebox+1pt\relax}%
    \ifdim\thealgruleheight<.75\baselineskip
        \def\thealgruleheight{\dimexpr .75\baselineskip+1pt\relax}%
    \fi
    \ifdim\thealgruledepth<.25\baselineskip
        \def\thealgruledepth{\dimexpr .25\baselineskip+1pt\relax}%
    \fi
    \State #1%
    \def\thealgruleheight{\dimexpr .75\baselineskip+1pt\relax}%
    \def\thealgruledepth{\dimexpr .25\baselineskip+1pt\relax}%
}

\newtheorem{theorem}{Theorem}

\newcommand{\abs}[1]{\left\vert#1\right\vert}
\newcommand{\set}[1]{\left\{#1\right\}}

\usepackage{mathtools}

\newcommand{\Cc}{\mathcal C}

\newcommand{\Gc}{\mathcal G}

\newcommand{\Lc}{\mathcal L}
\newcommand{\Mc}{\mathcal M}

\newcommand{\Pc}{\mathcal P}

\newcommand{\Sc}{\mathcal S}
\newcommand{\Rc}{\mathcal R}
\newcommand{\Tc}{\mathcal T}

\newcommand{\mv}{\mathbf m}

\newcommand{\yv}{\mathbf y}

\newcommand{\alg}{DAPP-ECC}
\newcommand{\migProb}{PMP}
\newcommand{\cpuall}{GFA}
\newcommand{\bu}{SFS} 

\newcommand{\pd}{PD}

\DeclareMathOperator{\pdMath}{\pd}

\newcommand{\pu}{PU}

\newcommand{\lBound}{LBound}
\newcommand{\notAssigned}{\tilde{\Mc}} 
\newcommand{\notAssignedOfS}{\notAssigned_s} 
\newcommand{\pushUpSet}{\Pc^U} 
\newcommand{\pushUpSetOfS}{\pushUpSet_s} 
\newcommand{\pushDownSet}{\Pc^D} 
\newcommand{\pushDownSetOfS}{\pushDownSet_s}

\newcommand{\deficitCpu}{deficitCpu}
\DeclareMathOperator{\deficitCpuMath}{\deficitCpu}

\newcommand{\accDelayBu}{T^{\bu}_{ad}}
\newcommand{\accDelayPd}{T^{\pd}_{ad}}

\newcommand{\ffit}{F-Fit}

\newcommand{\subFigWidth}{0.24 \textwidth}

\newcommand{\algFontSize}{\scriptsize}

\date{}
\title{
Asynchronous Distributed Protocol for Service Provisioning in the Edge-Cloud Continuum 
}
\author{
\IEEEauthorblockN{Itamar Cohen}
\IEEEauthorblockA{Ariel University\\ 
Ariel, Israel}
\and
\IEEEauthorblockN{Paolo Giaccone}
\IEEEauthorblockA{Politecnico di Torino\\ Torino, Italy}
\and
\IEEEauthorblockN{Carla Fabiana Chiasserini} 
\IEEEauthorblockA{Politecnico di Torino\\
Torino, Italy}
}

\begin{document}
\maketitle

\begin{abstract}
In the edge-cloud continuum, datacenters provide microservices (MSs) to mobile users, with each MS having specific 
latency constraints and computational requirements. Deploying such a variety of MSs matching their requirements with the available computing resources is challenging. In addition, time-critical MSs may have to be migrated as the users move, to keep meeting their latency constraints.  
Unlike previous work relying on a central orchestrator with an always-updated global view of the available resources and of the users' locations, this work envisions a distributed solution to the above issues. 
In particular, we propose a distributed asynchronous protocol for MS deployment in the cloud-edge continuum that (i) dramatically reduces the system overhead compared to a centralized approach, and (ii) increases the system stability by avoiding having a single point of failure as in the case of a central orchestrator. 
 Our solution ensures cost-efficient feasible placement of MSs, while using negligible bandwidth. 
\end{abstract}

\section{Introduction}
Today's networks offer bulk virtualized resources, embodied as a collection of datacenters on the continuum from the edge to the cloud~\cite{Justifies_path_to_root_n_CLP_vehs_short, SFC_mig_short,  orch_cloud2edge_survey_short, Justify_CLP_tree_ISPs}.
These datacenters host a plethora of applications with versatile computational requirements and latency constraints. For example, such time-critical services as road safety applications require low latency, which may dictate processing them in an edge datacenter, close to the user. In contrast, infotainment tasks require larger computational resources, but looser latency constraints and therefore may be placed on a cloud datacenter with abundant and affordable computation resources~\cite{tong2016_justify_path_to_root, SFC_mig_short}. 
Placing services over the cloud-edge continuum is thus challenging. It becomes even more complex when changes in the users' mobility or traffic demand require migrating services to reduce latency. 

Most of the existing solutions~\cite{Justifies_path_to_root_n_CLP_vehs_short, Dynamic_Service_Provisioning_ToN_short, SFC_mig_mechanism, orch_cloud2edge_survey_short, dynamic_sched_and_reconf_t_short, Avatar, SFC_mig_short} 
rely on a central orchestrator to make all placement and migration decisions.
The orchestrator periodically (i) gathers information about the state of resources and migration requirements, (ii) calculates new placement and resource allocation, and (iii) instructs datacenter local controllers accordingly.
This centralized synchronous approach has several shortcomings. First, it does not scale well, thus failing to manage systems with multiple datacenters efficiently. In practice, gathering fresh state information causes significant communication bottlenecks, even within a single cloud datacenter~\cite{APSR_short}.
Secondly, the orchestrator is a natural single point of failure, compromising the system's stability. Finally, the datacenters may be operated by distinct operators~\cite{Crosshaul, NFV_ego_learning}, which are typically unwilling to share proprietary information and implementation details with competitors. 

\emph{Our Contribution.} We present a solution, named Distributed Asynchronous Placement Protocol for the Edge-Cloud Continuum (\alg), that effectively and efficiently overcomes the above issues. 
\alg\ is carefully crafted to decrease communication overhead by using simple single-hop control messages transmitted by a node to only relevant neighbors.
Furthermore, \alg\ requires no out-of-band communication or synchronization tools.
\alg\ can find a feasible solution even with restricted resources, where a feasible placement necessitates migrating also already-placed MSs.
Finally, and very importantly, our solution allows multiple datacenters -- possibly of distinct providers -- to cooperate without exposing proprietary information. 

\emph{Paper organization.}
We introduce the system model in Sec.~\ref{sec:system} and formulate the placement and migration problem in Sec.~\ref{sec:problem}. Sec.~\ref{sec:alg} describes our algorithmic solution. Sec.~\ref{sec:sim} evaluates the performance of \alg\ in various settings, using real-world mobility traces and antenna locations. Finally,  Sec.~\ref{sec:related_work} reviews relevant related work, and Sec.~\ref{sec:conclusion} draws some conclusions.

\section{System Model}\label{sec:system}

We consider a fat-tree cloud-edge continuum  architecture, which comprises~\cite{tong2016_justify_path_to_root}: 
\begin{inparaenum}[(i)]
\item a set of {\em datacenters}, $\Sc$, denoting generic computing resources,
\item {\em switches}, and
\item {\em radio Points of Access (PoA)}.
\end{inparaenum}
Datacenters are connected through switches, and PoAs have a co-located datacenter~\cite{Mig_in_Mobile_Edge_Clouds_short}. 
Each user is connected to the network through a PoA, and they can change PoA as they move.

 We model such logical multi-tier network as a directed graph ${\Gc} = (\Sc,\Lc)$, where   
the vertices are the datacenters, while the edges are the directed virtual links connecting them.
We assume the existence of a single predetermined loop-free path between each pair of datacenters. 

Let us consider a generic user generating a {\em service request} $r$, originating at the PoA $p^r$, to which the user is currently connected. 
Each request is served by placing an instance of a microservice (MS) on a datacenter.
Denote the instance of the MS for service request $r$ by $\mv^r$.
Let $\Rc$ denote the set of service requests, and $\Mc$ the set of corresponding MSs that are currently placed, or need to be placed, on datacenters.

Each service is associated with an SLA, which specifies its requirements in terms of KPI target values~\cite{OKPI_short}.
Let us consider {\em latency} as the most relevant KPI, although our model could be extended to others, like throughput and energy consumption.
Due to these latency constraints, each request $r$ is associated with a list of {\em delay-feasible} datacenters $\Sc_r$. 
The delay-feasible servers in $\Sc_r$ are not too far from $r$'s PoA ($p^r$), or, more formally, their list is a prefix of the path from  $p^r$ to the root~\cite{Justifies_path_to_root_n_CLP_vehs_short, Justify_CLP_tree_ISPs, Dynamic_Service_Provisioning_ToN_short}.
The top delay-feasible datacenter of request $r$ is denoted by $\bar{s}_r$. 

To successfully serve request $r$ on datacenter $s \in \Sc_r$, $s$ should allocate (at least) $\beta^{r,s}$ CPU units, where $\beta^{r,s}$ is an integer multiple of a basic CPU speed. 
As there exists a known method to calculate $\Sc_r$ and $\beta^{r,s}$ given 
the characteristics of $\mv^r$~\cite{Dynamic_Service_Provisioning_ToN_short}, we refer to $\Sc_r$ and $\beta^{r,s}$ as known input parameters. 
Each datacenter $s\in \Sc$ has a total processing capacity $C_s$, expressed in number of CPU cycles/s.

\section{The Placement and Migration  Problem}\label{sec:problem}

The delay experienced by an MS may vary over time due to either
\begin{inparaenum}[(i)]
\item a change in the user's PoA, which changes the network delay, or
\item a fluctuation in the traffic, and hence in the processing latency~\cite{Dynamic_user_demands}.
\end{inparaenum}
Each user continuously monitors its Quality of Experience (QoE) and warns its PoA as its latency approaches the maximum acceptable value\footnote{If the user can predict its near-future location, it can inform the PoA before the target delay is violated.}. 
The PoA then checks the request, and if the user is indeed {\em critical} -- namely, its latency constraint is about to be violated -- the PoA triggers a migration algorithm. The PoA also handles new requests that are yet to be served.

{\bf Decision variables.} 
Let $\yv$ be the Boolean placement decision variables, i.e., $y(r,s) = 1$ if MS $\mv^r$ is scheduled to run on datacenter $s$.
Any choice for the values of such variables provides a {\em solution} to the {\em Placement and Migration Problem (PMP)}, determining (i) where to  deploy new MSs, (ii) which existing MSs to migrate, and (iii) where to migrate them.

{\bf Constraints.} 
The following constraints hold:
\begin{align}
\textstyle
\sum_{s \in \Sc} y(r,s)
&= 1
&\forall r \in \Rc \label{problem_constraint:single_placement} \\
\textstyle
\sum_{r \in \Rc} y(r,s) \beta^{r,s}
&\leq C_s
&\forall s \in \Sc\,. \label{problem_constraint:datacenter_computational_residual_capacity}
\end{align}
Constraint \eqref{problem_constraint:single_placement} ensures that at any point in time, each MS $\mv^r$ is associated with a single {\em scheduled} placement. \eqref{problem_constraint:datacenter_computational_residual_capacity} assures that the capacity of each datacenter is not exceeded.

{\bf Costs.}
The system costs are due to migration and computational resource usage, as detailed below.

Migrating MS $\mv^r$ from datacenter $s$ to datacenter $s'$ incurs a {\em migration cost}   $\psi^m(r,s,s')$. 
Let $x(r,s)$ denote the {\em current placement indicator parameters}\footnote{$x(r,s)$ are not decision variables, as they indicate the current deployment.}, i.e., $x(r,s) = 1$ iff MS $\mv^r$ is currently placed on datacenter $s$. We assume that a user does not become critical again before it finishes being placed based on the decision made by any previous run of the algorithm solving the PMP. 
The migration cost incurred by a critical MS $\mv^r$ is then:
\begin{align}
\sum_{s\neq s' \in \Sc} x(r,s) \cdot y(r,s') \cdot \psi^m(r,s,s'). \notag
\end{align}

Placing MS $r$ on datacenter $s$ incurs a {\em computational cost} $\psi^c(r,s)$. As computation resources in the cloud are cheaper~\cite{tong2016_justify_path_to_root, SFC_mig_short}, we assume that if $s$ is an ancestor of $s'$, placing MS $\mv^r$ on $s$ is cheaper than placing $\mv^r$ on $s'$.

{\bf Objective.}
Our goal is to minimize the cost function:
\begin{align}\label{Eq:def_obj_func}
\phi(\yv) &=
\sum_{{s\neq s' \in \Sc }} 
\sum_{r \in \Cc} 
x(r,s) \cdot y(r,s') \cdot \psi^m(r,s,s') \\ \notag
&+ 
\sum_{s \in \Sc} y(r,s) 
\sum_{r \in \Cc} 
\psi^c(r,s)
\end{align}
subject to constraints~\eqref{problem_constraint:single_placement},
\eqref{problem_constraint:datacenter_computational_residual_capacity}.

Following Proposition\,2 in~\cite{Dynamic_Service_Provisioning_ToN_short}, it is easy to see that \migProb\ is NP-hard.
We are interested in a \textit{distributed} solution, where no single datacenter (or any other entity) has a complete fresh view of the status (e.g., the current place of each MS, or the amount of available resources in each datacenter). Instead, the  placement and migration protocol should run on an as small as possible subset of the datacenters.
Furthermore, the solution should be \textit{asynchronous}, as distinct PoAs may independently invoke different, simultaneous runs of the protocol.

\section{The \alg\ Algorithmic Framework}\label{sec:alg}

In this section, we present our algorithmic solution to \migProb, named  Distributed Placement Protocol for the Edge-Cloud Continuum (\alg). We  start with a high-level description and then provide the details of the single algorithms. 
In our description, we let $s$.proc() denote a run of procedure proc() on datacenter $s$. As our protocol is distributed, each datacenter $s$ maintains its local variables, denoted by a sub-script $s$. 
We will use the procedure {\em Sort()} that sorts MSs in non-increasing timing criticality, realized by a non-decreasing $\abs{\Sc_r \setminus \Tc(s)}$, i.e., the number of ancestor datacenters on which the MS may be placed. {\em Sort()} breaks ties by non-decreasing $\beta^{r,s}$ and breaks further ties by users' FIFO order. 

\subsection{Protocol overview}\label{sec:alg:hi-lvl}
Following the intuition, one would reduce the system costs by placing MSs in the network continuum as close as possible to the cloud, since cloud resources are  cheaper and this may prevent future migrations. However, such an approach may make the algorithm fail to find feasible solutions, even when they exist~\cite{Dynamic_Service_Provisioning_ToN_short}. 

Our solution to this conflict between feasibility and cost-efficiency is inspired to~\cite{Dynamic_Service_Provisioning_ToN_short}. The proposed \alg\ algorithm initially {\em assigns} -- or, better, reserves -- CPU for each request {\em as close as possible to the edge}. We dub this stage Seek a Feasible Solution (SFS). Once such a solution is found, the protocol Pushes Up (PU) the  MSs as much as possible {\em towards the cloud}, to reduce costs. 
If SFS cannot find a feasible solution,  non-critical MSs will be migrated via the Push-Down (PD) procedure, to make room for a critical MS.

\subsection{The \alg\ algorithms}
We now detail the algorithmic framework we developed. We will denote by $\notAssigned$ a list of currently unassigned requests, and by $\pushUpSet$ a list of assigned requests that may be pushed-up to a closer-to-the-cloud datacenter, to reduce costs. Let $\pushDownSet$ denote a set of push-down requests.
$a_s$ denotes the available capacity on datacenter $s$. Upon system initialization, each datacenter $s$ assigns $\notAssignedOfS=\pushUpSetOfS=\pushDownSetOfS=\emptyset$ and $a_s=C_s$.

\subsubsection*{Seek for a feasible solution} $s$.\bu() is presented in Alg.~\ref{alg:bu}. 
It handles the unassigned MSs as follows. If the locally available capacity suffices to locally place an unassigned MS $\mv^r$ (Ln.~\ref{alg:bu:if_enough_avail_CPU}), $s$ reserves capacity for $\mv^r$ (Ln.~\ref{alg:bu:dec_a_s}). If $\mv^r$ cannot be placed higher in the tree (Ln.~\ref{alg:bu:if_must_place}), $s$.\bu() not only assigns $\mv^r$, but also locally places it. Otherwise, the procedure inserts $\mv^r$ to the set of potentially-placed MSs, which $s$ will later propagate to its parent. If $s$.\bu() fails to place a request $r$ that cannot be placed higher, it calls $s$.\pd() (Lines~\ref{alg:bu:if_failed_begin}-\ref{alg:bu:if_failed_end}). 
The arguments for $s$.\pd() are 
(i) the identity of the initiator datacenter, $s^*$; (ii) a list of MSs that $s^*$ asks its descendants to push-down, 
and (iii) deficitCPU, namely, the amount of CPU resources that must be freed from $s^*$ to find a feasible solution.
    
In Lines~\ref{alg:bu:set_puList_to_prnt}-\ref{alg:bu:if_should_call_prnt}, $s$.\bu() checks whether there exist MSs that are not yet assigned, or may be pushed-up to an ancestor. If so, $s$.\bu() initiates a run of \bu() on $s$'s parent (Ln.~\ref{alg:bu:call_prnt}). 
If there are no pending push-up requests from its ancestors, $s$ initiates a push-up (Lines~\ref{alg:bu:if_no_pnding_pu_from_prnt}-\ref{alg:bu:init_pu}).

\begin{algorithm}
\caption{$s$. \bu ($ \notAssigned, \pushUpSet$)}
 \algFontSize
\label{alg:bu}
\algFontSize
\begin{algorithmic}[1]
    \State $\pushUpSetOfS   \leftarrow \pushUpSetOfS \cup \pushUpSet$
    \State $\notAssignedOfS \leftarrow$ Sort ($\notAssignedOfS \cup \notAssigned$)  
    \label{alg:bu:sort}
    \ForEach {$\mv^r \in \notAssignedOfS$}
        \label{alg:bu:for_begin}
        \If {$a_s \geq \beta^{r,s}$}
        \Comment{enough available CPU to place $\mv^r$ on $s$}
        \label{alg:bu:if_enough_avail_CPU}
        \State remove $\mv^r$ from $\notAssignedOfS$
        \label{alg:bu:pup_from_Hcomp}
        \State $a_s \leftarrow a_s - \beta^{r,s}$        
        \Comment{assign $\mv^r$ on $s$}
        \label{alg:bu:dec_a_s}
        \If {$\bar{s}_r = s$} 
        \Comment{must place $\mv^r$ on $s$ for a feasible sol}
          \label{alg:bu:if_must_place}
            \State place $\mv^r$ on $s$
            \label{alg:bu:place}
        \Else
        \Comment{Will place $\mv^r$ on $s$ only if it won't be pushed-up}
            \State insert $\mv^r$ to $s$.potentiallyPlacedRequests\label{alg:bu:mark}
            \State insert $(\mv^r, s)$ to $\pushUpSetOfS$
            \label{alg:bu:insert_to_push_up_list}
        \EndIf
        \ElsIf {$\bar{s}_r = s$} 
            \label{alg:bu:if_failed_begin}
            \Comment{$\mv^r$ can't be placed here, nor higher in the tree}
            \State $\pushDownSetOfS \leftarrow \set{r \ | \ \mv^r \in \notAssignedOfS \ \AND \ \bar{s}(r)=s}$
            \label{alg:bu:set_pdList}
            \Comment{''over-provisioned'' MSs}
            \State $\deficitCpuMath = \sum_{r \in \pushDownSetOfS} \beta^{r,s} - a_s$
            \label{alg:bu:set_defCpu}
            \Comment{capacity to free for finding a sol}
            \State run $\pdMath\ (\pushDownSetOfS, \deficitCpuMath, s)$ 
            \label{alg:bu:call_pd}
        \EndIf
        \label{alg:bu:if_failed_end}
    \EndFor 
    \label{alg:bu:for_end}
    \State $\pushUpSet_{prnt} \leftarrow \set{(\mv^r,s') | (\mv^r,s') \in \pushUpSetOfS \ \AND \ s.\textrm{prnt} \in \Sc_r}$
    \label{alg:bu:set_puList_to_prnt}
    \If {$\notAssignedOfS \neq \emptyset$ \OR $\pushUpSet_{prnt} \neq \emptyset$}
    \Comment{is there any req to send to parent?}
    \label{alg:bu:if_should_call_prnt}
        \State{send \Big($s$.parent, \bu \big($\notAssignedOfS, \pushUpSet_{prnt}$\big)\Big)}
        \label{alg:bu:call_prnt}
        \State $\pushUpSet_s = \pushUpSet_s \setminus \pushUpSet_{prnt}$
    \EndIf        
    \If {$\pushUpSet_{prnt} = \emptyset$} 
    \Comment{No pending PU replies from parent}
    \label{alg:bu:if_no_pnding_pu_from_prnt}
        \State run $s$.\pu $\left(\pushUpSetOfS \right)$
        \label{alg:bu:init_pu}
    \EndIf
\end{algorithmic}
\end{algorithm}

\begin{algorithm}
\caption{$s$.\pu ($\pushUpSetOfS$)}
 \algFontSize
\label{alg:pu}
\begin{algorithmic}[1]
    \State dis-place all MSs pushed-up from me, and update $a_s$ and $\pushUpSetOfS$ accordingly
    \label{alg:pu:displace}
    \State $\pushUpSetOfS \leftarrow$ Sort ($\pushUpSetOfS$)
    \label{alg:pu:sort}
    \ForEach {$(\mv^r, s') \in \pushUpSetOfS$}
        \label{alg:pu:for_pu_begin}
        \If {$a_s \geq \beta^{r,s}$}
        \label{alg:pu:enough_as}
        \State {place $\mv^r$ on $s$ and decrease $a_s$ accordingly}
        \label{alg:pu:pu}
            \State remove $(\mv^r, s')$ from $\pushUpSetOfS$ and insert $(\mv^r,s)$ into $\pushUpSetOfS$ 
        \label{alg:pu:inform}
            \EndIf
    \EndFor    
    \label{alg:pu:for_pu_end}
    
    \ForEach{child $c$}\label{alg:pu:for_each_child_begin} 
        \State $\pushUpSet_c = \set{(\mv^r, s') \ | \ (\mv^r, s') \in \pushUpSetOfS \ \AND \ c \in \Sc_r}$
        \label{alg:pu:associate_puReq_to_child}
        \If{$\pushUpSet_c \neq \emptyset$}
            \label{alg:pu:forEach_child_if_begin}
            \State send \Big($c$, \pu \big($\pushUpSet_c\big)\Big)$
            \State $\pushUpSetOfS = \pushUpSetOfS \setminus \pushUpSet_c$
        \EndIf
        \label{alg:pu:forEach_child_if_end}
    \EndFor\label{alg:pu:for_each_child_end}
\end{algorithmic}
\end{algorithm}

\begin{algorithm}
\caption{
Datacenter $s$ called by \pd ($\pushDownSetOfS, \deficitCpuMath, s^*$)}
\algFontSize
\label{alg:pd}
\begin{algorithmic}[1]
    \If {$s_s^* \neq$ null \AND $s_s^* \neq s^*$}\Comment{Running another push-down procedure}
        \label{alg:pd:if_within_another_begin}
        \State send (caller, \pd ($\pushDownSetOfS$, deficitCpu, $s^*$))
        \State return
    \EndIf
    \label{alg:pd:if_within_another_end}
    \State $s_s^* \leftarrow s^*$
    \label{alg:pd:update_local_initiator}
    \State $\notAssignedOfS \leftarrow$ Sort ($\notAssignedOfS$)
    \State add $s\text{.potPlacedRequests}$, and then $s\text{.placedRequests}$ to the end of
    $\pushDownSetOfS$
    \label{alg:pd:add2pdList}
    \ForEach {child $c$} 
        \label{alg:pd:foreach_child_begin}
        \If {I can push-down to myself enough MSs from $s^*$ to nullify deficitCpu}
        \label{alg:pd:if_can_break_next_child_begin}
            \State break \label{alg:pd:if_can_break_next_child}
        \EndIf
        \label{alg:pd:if_can_break_next_child_end}
        \State $\pushDownSet_c = \set{\mv^r \ | \ r \in \pushDownSetOfS, c \in \Sc_r}$
        \Comment{MSs to push-down relevant to child $c$}
        \If {$\pushDownSet_c \neq \emptyset$}
            \label{alg:pd:call_child_begin}
            \Comment{child $c$ may help in freeing space}
            \State $\pushDownSet_c, \deficitCpuMath = c.\pdMath$  ($\pushDownSet_c, \deficitCpuMath, s^*$)
            \Comment{req. $c$ \& get ack}
            \label{alg:async_PrepResh:call_child}
            \State dis-place MSs pushed-down from me; update $a_s$,  $\pushDownSetOfS, \notAssignedOfS$ accordingly
        \EndIf
        \label{alg:pd:call_child_end}
    \EndFor
    \label{alg:pd:foreach_child_end}
    \ForEach {$r \in \pushDownSetOfS$ s.t. $r$.curPlace is $s$ or an ancestor of $s$ \AND $\beta^{r,s} \leq a_s$}
    \label{alg:pd:pd_to_myself_begin}
        \State {place $\mv^r$ and update $a_s$, $\pushDownSet_c$ and deficitCpu accordingly}
        \label{alg:pd:pd_to_myself_end}
    \EndFor
    \If {$s \neq s_s^*$} 
    \Comment{I'm not the initiator of this reshuffle}
    \label{alg:pd:if_need_callPrnt}
        \State send \Big($s$.parent ($\pushDownSetOfS, \deficitCpuMath, s^*_s$)\Big)
        \label{alg:pd:callPrnt}
    \EndIf
    \State run $s$.\bu($\notAssignedOfS$) in $F$-mode
    \State $s_s^* \leftarrow$ null
    \end{algorithmic}
\end{algorithm}

\subsubsection*{Push-Up}   $s$.\pu(), detailed in Alg.~\ref{alg:pu},  first displaces and regains the CPU resources for all the MSs pushed-up from $s$ to a higher-level datacenter.
Next, $s$.\pu() handles all the push-up requests as follows. Consider a request to push up MS $\mv^r$, currently placed on datacenter $s'$ that is a descendent of $s$. If $s$ has enough available capacity for that request, then $s$.\pu() locally places $\mv^r$ (Lines~\ref{alg:pu:enough_as}-\ref{alg:pu:pu}) and updates the relevant record in $\pushUpSetOfS$ (Ln.~\ref{alg:pu:inform}). This record will later be propagated to $s'$ which will identify that $\mv^r$ was pushed up, and regain the resources allocated for it.
In Lines~\ref{alg:pu:for_each_child_begin}-\ref{alg:pu:for_each_child_end}, $s$ propagates the push-up requests to its children. To reduce communication overhead, each push-up request in $\pushUpSetOfS$ is propagated only to the child $c$ that is delay-feasible for the MS in question.

\subsubsection*{Push-Down} $s$.\pd(), in Alg.~\ref{alg:pd}, runs the same when either a parent calls its child, or vice versa. $s^*_s$ denotes the initiator of the \pd\ procedure currently handled by datacenter $s$. If no \pd\ procedure is currently handled by  $s$, then $s^*$=null. Note that several parallel simultaneous runs of \pd\ may exist in the system. Each such run is unequivocally identified by its initiator $s^*$. At any time instant, each such run is associated with a single current value of deficitCpu.
 \pd\ runs sequentially, in a DFS manner, in the sub-tree rooted by $s^*$, and terminates once deficitCPU is nullified.

If $s$.\pd() is invoked while $s$ takes part in another run of \pd() (realized by a different initiator $s^*$), the procedure replies with the minimal data necessary to retain liveness (Lines~\ref{alg:pd:if_within_another_begin}-\ref{alg:pd:if_within_another_end}). Otherwise, $s$.\pd() adds to the given set of requests $\pushDownSetOfS$ its locally assigned MSs. To reduce the number of migrations, the locally assigned MSs are added to the end of $\pushDownSetOfS$, so that the procedure will migrate already-placed MSs only if necessary for finding a feasible solution.
In Lines~\ref{alg:pd:foreach_child_begin}-\ref{alg:pd:foreach_child_end}, $s$ serially requests its children to push-down MSs, to free space in $s^*$. The amount of space to be freed from $s^*$ is \deficitCpu. 
Before each such call, $s$ checks whether nullifying \deficitCpu\ without calling an additional child is possible. If the answer is positive, $s$ skips calling its children (Lines~\ref{alg:pd:if_can_break_next_child_begin}-\ref{alg:pd:if_can_break_next_child_end}). Upon receiving a reply from child $c$, the procedure updates \deficitCpu\ and $s$'s state variables according to the MSs that were pushed-down to $c$'s sub-tree (Lines~\ref{alg:pd:call_child_begin}-\ref{alg:pd:call_child_end}).
In Lines~\ref{alg:pd:pd_to_myself_begin}-\ref{alg:pd:pd_to_myself_end}, $s$.\pd() tries to push-down to $s$ MSs from the push-down list, $\pushDownSetOfS$. 
Later, if $s$ is not the initiator of this push-down procedure, it calls its parent (Lines~\ref{alg:pd:if_need_callPrnt}-\ref{alg:pd:callPrnt}).
Finally, $s$.\pd() calls \bu\ in $F$-mode (described below) to place all its yet-unassigned MSs, if such exist.

The following theorem assures the convergence of \alg ~(proof omitted due to space constraints). 
\begin{theorem}\label{thm:converge}
If there are no new requests, the protocol either fails or finds a feasible solution after exchanging a finite number of messages.
\end{theorem}

\subsection{Reducing the communication overhead}\label{sec:practical_imp}

\textbf{F-mode.}
Intuitively, a run of $s$.\pd() indicates that a recent run of \bu() -- either in $s$, or in an ancestor of $s$ -- failed, and hence called  \pd(). Since in such a case, there is a high risk of failing again, the algorithm should focus on finding a feasible solution, rather than reducing costs: it does not make sense to push-up MSs just to push them back down slightly later.
Hence, we define an {\em F} (feasibility)-mode of the protocol. Each time $s$.\pd() is called, $s$ enters $F$-mode (if it was not already in $F$-mode), and remains so for some pre-configured {\em F-mode period}.
While in $F$-mode, \alg\ does not initiate new push-up requests, and only replies to existing push-up requests with the minimum necessary details to prevent deadlocks. If $s$.\bu() does not find a feasible solution while $s$ is in $F$-mode, \alg\ terminates with a failure.

\textbf{Accumulation delay.}
Theoretically, each attempt to place a single MS may result in a unique run of  \pd\  that involves all the datacenters, thus incurring excessive overhead. To avoid such case, 
observe that, typically, several users move together in the same direction (e.g.,  cars moving simultaneously on the road, on the same trajectory). 
Naively, such a scenario may translate to multiple invocations of \alg, each of them for placing a single request. 
To tackle this problem, we 
introduce short {\em accumulation delays} to our protocol. 
We let each datacenter receiving a \bu\ message wait for a short {\em \bu\ accumulation delay} before it begins handling the new request. 
To deter long service delays or even deadlocks, each datacenter maintains a single \bu\ accumulation delay timer that operates as follows: if a run of \bu\ reaches Ln.~\ref{alg:bu:for_begin} in Alg.~\ref{alg:bu} while no \bu\ accumulation delay timer is ticking, the procedure initiates a new \bu\ accumulation delay timer. This current run of \bu, as well as all the subsequent runs, halt. After the \bu\ accumulation delay terminates, only a single \bu\ process resumes (see Alg.~\ref{alg:bu}, Ln.~\ref{alg:bu:for_begin}).

Likewise, to initiate fewer runs of \pd(), we let each datacenter retain a single \pd\ accumulation delay mechanism that works similarly to the \bu\ accumulation delay timer. 
Significantly, the accumulation delay only impacts the time until the protocol finds a new feasible placement, not the delay experienced by applications in the data plane. We assess the impact of the accumulation delay in Sec.~\ref{sec:sim:feasibility}. 

\section{Numerical Evaluation}\label{sec:sim}

\subsection{Simulation settings}\label{sec:sim_settings}

\textbf{Service area, network, and datacenters.}
We consider two mobility traces, representing real-world scenarios with distinct characteristics: the vehicular traffic within the centers of the cities of (i) Luxembourg~\cite{Luxembourg_short},  and (ii) the  Principality of Monaco~\cite{Monaco}.   
For the PoAs, we rely on real-world antenna locations, publicly available in~\cite{OpenCellid}. For each simulated area, we consider the antennas of the cellular telecom provider having the largest number of antennas in the simulated area.
For both traces, we consider the 8:20-8:30 am rush hour period. 
Further details about the mobility traces can be found 
in~\cite{Dynamic_Service_Provisioning_ToN_short, Luxembourg_short, Monaco}.


\textbf{Network and datacenters.}
The cloud-edge continuum is structure as a 6-height tree;  a topology level is denoted by $\ell\in\{0, 1, \dots, 5\}$. The leaves (level 0) are the datacenters co-located with the PoAs (antennas). 
Similarly to~\cite{Avatar, dynamic_sched_and_reconf_t_short, NFV_ego_learning}, the higher levels $5, 4, 3, 2, 1$ recursively partition the simulated area. 
In both Luxembourg and Monaco, if no PoAs exist in a particular rectangle, the respective datacenters are pruned from the tree. 
The CPU capacity increases with the level $\ell$ to reflect the larger computational capacity in datacenters closer to the cloud. Denoting the CPU capacity at each leaf datacenter by $C_\text{cpu}$, the CPU capacity in level $\ell$ is $(\ell+1) \cdot C_\text{cpu}$. 

\textbf{Services and costs.} 
Each vehicle that enters the considered geographical area is randomly marked as requesting either real time (RT) or non-RT services, with some probability defined later.
We calculate $\Sc_r$, $\beta^{r,s}$, and $\psi^c(r,s)$ for each $r \in \Rc, s \in \Sc$ using the \cpuall\ algorithm and the same data-plane latency parameters as in~\cite{Dynamic_Service_Provisioning_ToN_short}.
We thus obtain the following values. 
Each RT request can be placed on levels 0, 1, or 2 in the tree, requiring CPU of 17, 17, and 19~GHz, associated with costs of 544, 278, and 164, respectively. 	  
Each non-RT request can be placed on any level, with a fixed allocated CPU of 17~GHz and associated costs of 544, 278, 148, 86, 58, and 47 for placing the MS on levels 0, 1, 2, 3, 4 and 5, respectively. 
The migration cost is $\psi^m(r,s,s') = 600$ for every request $r$ and datacenters $s, s'$.

\textbf{Delays.} 
The delay experienced by each packet consists of (i)  transmission delay and (ii) propagation delay.

The transmission delay is calculated as the packet's size over the capacity allocated for the control plane at each link, through a dedicated network slice. We assume that this capacity is 10~Mbps. 
We now detail the size of each field in the messages exchanged by \alg. As \alg\ uses only single-hop packets, we assume a fixed 80-bits header. The IDs of datacenters, and requests, are represented using 12-bits, and 14-bits. Each MS belongs to a concrete {\em class} of timing constraint, expressed through a 4-bit classId. 
The CPU allocation of an MS on a datacenter $\beta^{r,s}$ is represented through a 5-bits field. \deficitCpu\ is at most the highest capacity of any single datacenter; we assume that this requires 16 bits. 
For the propagation delay, we use a pessimistic approach, where the length of every single link in the network corresponds to the diameter of the simulated area, and the propagation speed is $2 \cdot 10^8~\text{ms}^{-1}$. Consequently, the propagation delay of each link in Luxembourg and Monaco is $22~\mu \text{s}$ and $8~\mu \text{s}$ (resp.). 
For a datacenter at level $\ell$,  \bu\ accumulation delay, and \pd\ accumulation delay are 
$(\ell+1) \cdot \accDelayBu$, 
and $(\ell+1) \cdot \accDelayPd$ (resp.). 
We assign $\accDelayBu=0.1$\,ms and $\accDelayPd=0.4$\,ms.
$F$-mode period (recall Sec.~\ref{sec:practical_imp}) is 10\,s.

\textbf{Benchmark algorithms.} 
We are unaware of any fully distributed, asynchronous algorithm for the \migProb. Hence, we consider {\em centralized} placement schemes that identify the currently critical and new users once in a second and solve the respective  \migProb. We will consider the following algorithms.

{\em Lower Bound (LBound):} An optimal solution to the \migProb\ that can place fractions of an MS on distinct datacenters. Also, the LP formulation considers all MSs in the system every 1~s period, not just critical MSs. Hence, it serves as a lower bound on the cost of any feasible solution to the problem.

{\em \ffit}: It places each request $r$ on the lowest datacenter in $\Sc_r$ that has sufficient available resources to place MS $\mv^r$. This is an adaptation to our problem of the placement algorithm proposed in Sec.\,IV.B  in~\cite{App_placement_in_fog_n_edge}. 

{\em BUPU~\cite{Dynamic_Service_Provisioning_ToN_short}:} It consists of two stages. At the {\em bottom-up}, it places all the critical and new MSs as low as possible. If this stage fails to place an MS while considering datacenter $s$, the algorithm re-places {\em all} the MSs associated with $s$'s sub-tree from scratch. 
Later, BUPU performs a push-up stage similar to our \alg's \pu() procedure. 



\textbf{Simulation methodology.}
We simulate users' mobility using SUMO~\cite{SUMO2018_short}. The benchmark algorithms use the Python code publicly available in~\cite{SFC_mig_Github}. 
\alg\ is implemented using OMNeT++ network simulator~\cite{omnet}. Each new user, or an existing user which becomes critical, invokes a run of \alg\ datacenter co-located with the user's PoA. \alg's code is available in~\cite{dist_SFC_mig_Github}.
LBound is computed using Gurobi optimizer~\cite{Gurobi}.

\subsection{Resources required for finding a feasible solution}\label{sec:sim:feasibility}

\begin{figure}
    \centering
    \subfloat[\label{fig:cpu_vs_RT_prob_Lux}Luxembourg] {
        \includegraphics[width=\subFigWidth]{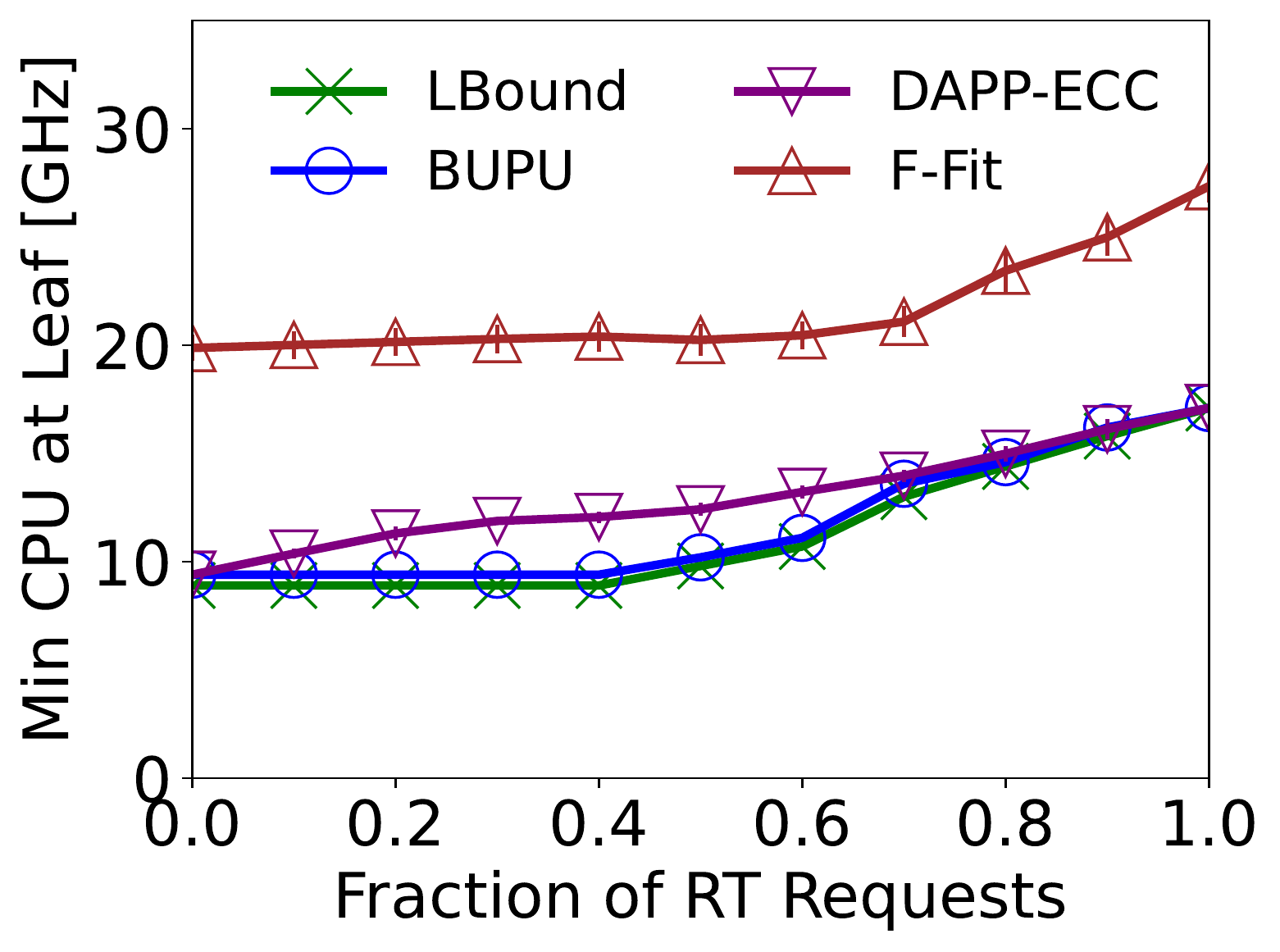}
    }
    \subfloat[\label{fig:cpu_vs_RT_prob_Monaco}Monaco] {
    \includegraphics[width=\subFigWidth]{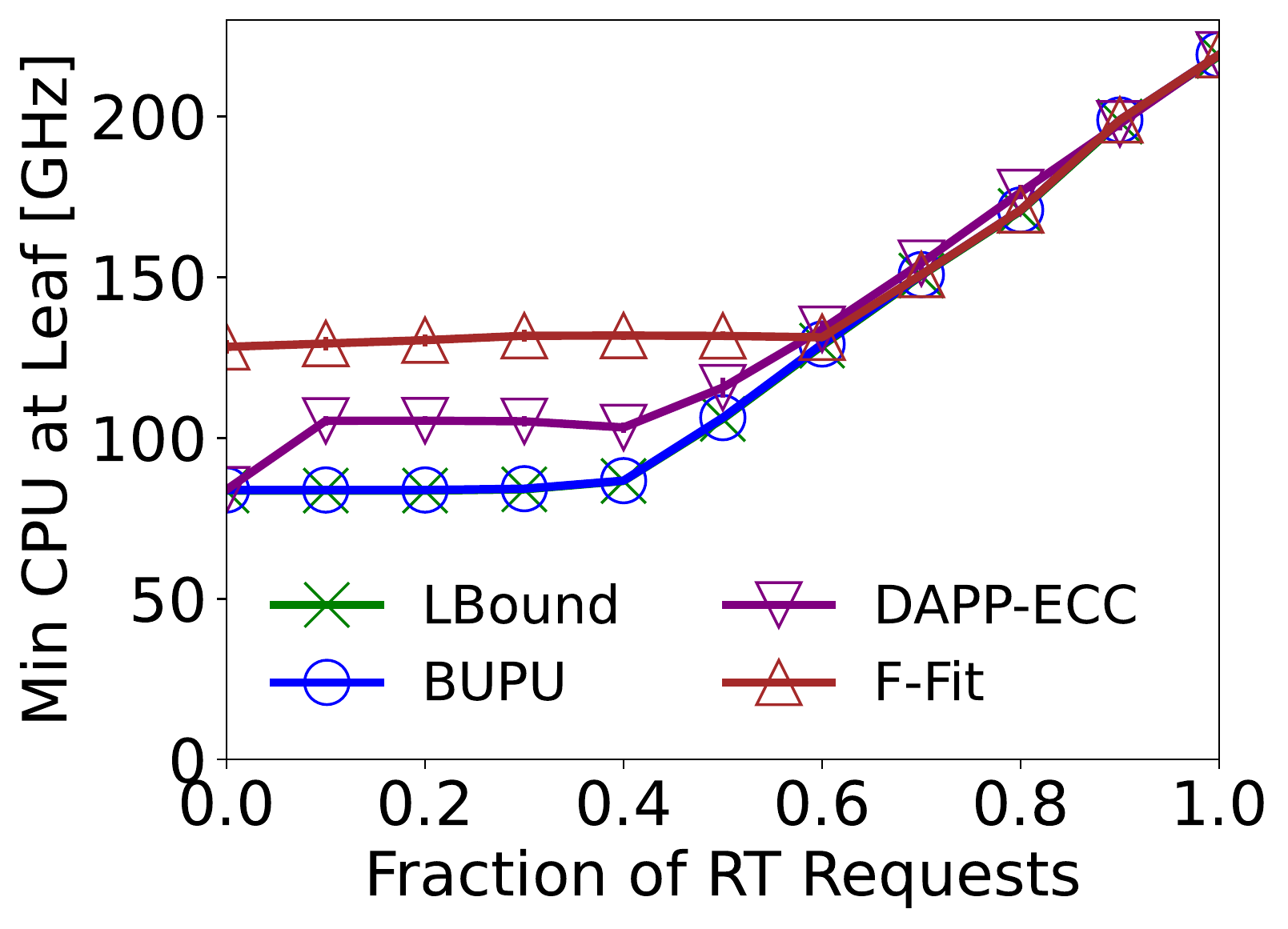}
    }
    \caption{Minimum required processing capacity for finding a feasible solution when varying the ratio of RT service requests.}
    \label{fig:cpu_vs_RT_prob}
\end{figure}

We now study the amount of resources
each algorithm requires to find a feasible solution. 
We vary the fraction of RT MSs with respect to the total MSs.
For each setting, a binary search is used to find the minimum amount of resources needed by the algorithm in question to successfully place all the critical MSs along the trace.
Fig.~\ref{fig:cpu_vs_RT_prob} presents the results of this experiment. The amount of CPU required by BUPU is almost identical to the optimal value. Despite being fully distributed and asynchronous, the amount of CPU needed by \alg\ is only slightly higher than BUPU. 
Finally, \ffit\ requires a processing capacity that is 50\% to 100\% higher than \lBound\ to provide a feasible solution.

\begin{figure}
    \centering
    \subfloat[\label{fig:ComOh_Lux}Luxembourg] {
        \includegraphics[width=\subFigWidth]{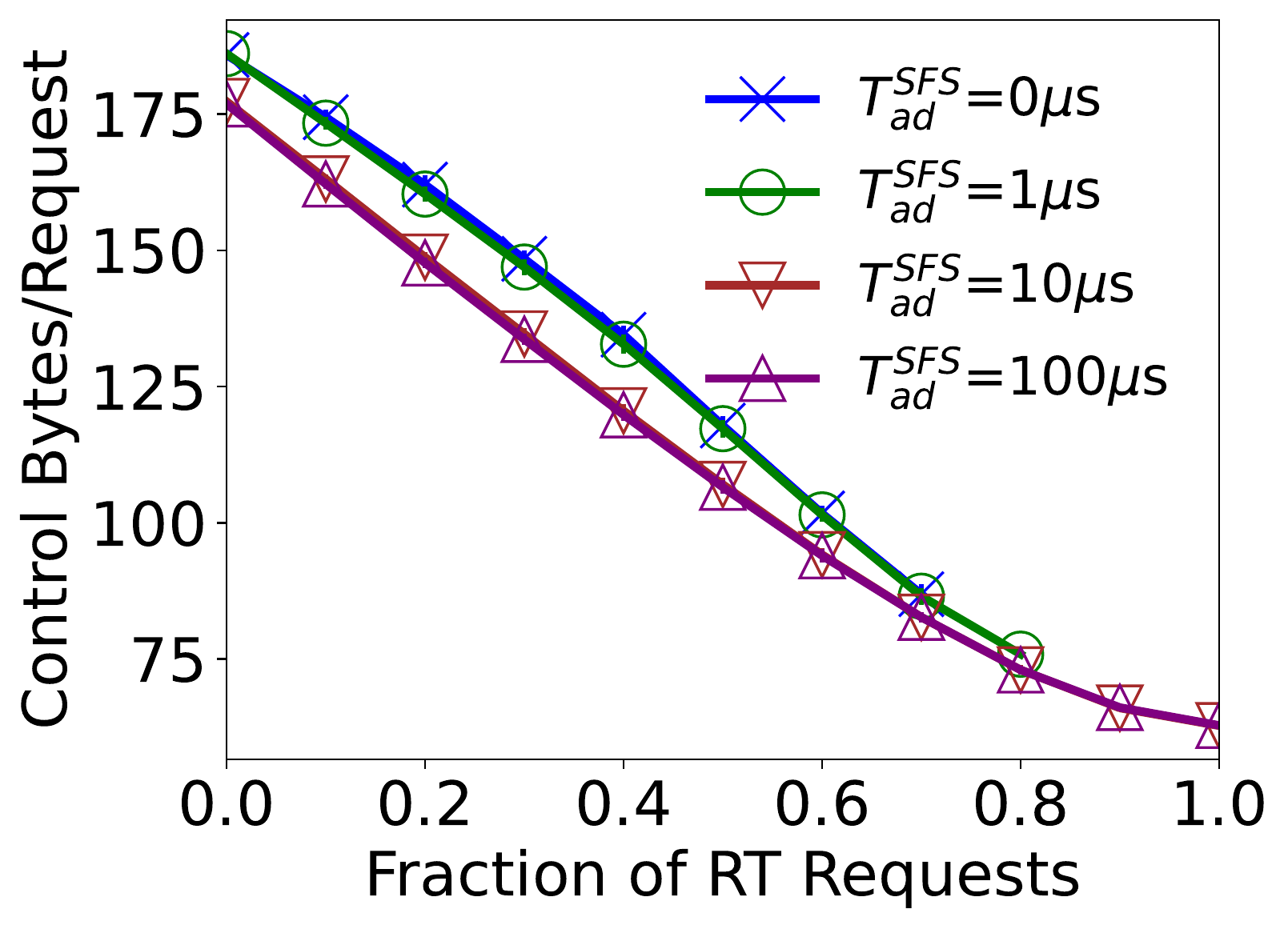}
    }
    \subfloat[\label{fig:ComOh_Monaco}Monaco] {
        \includegraphics[width=\subFigWidth]{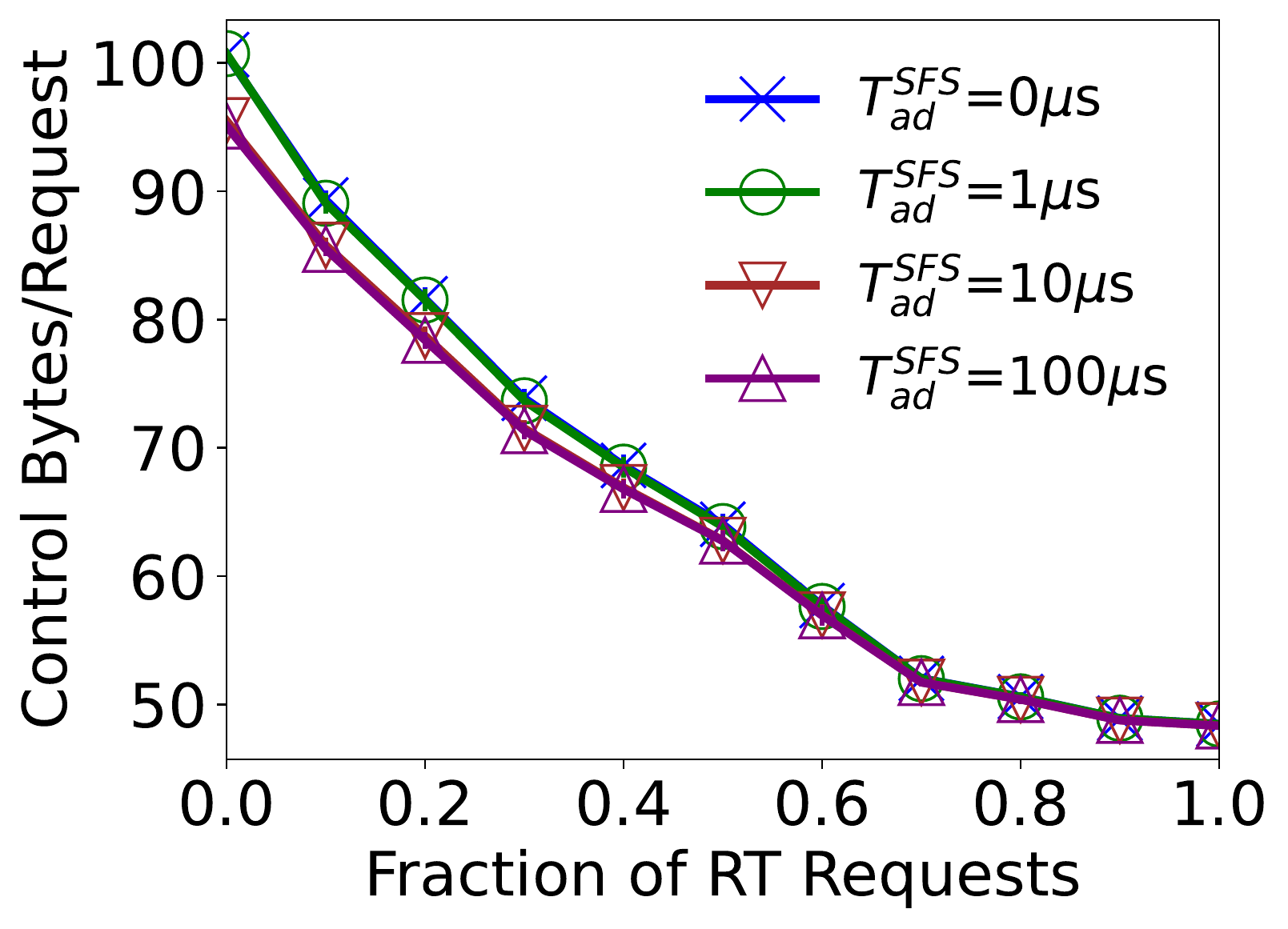}
    }
    \caption{Per-request signaling overhead due to \alg.}
    \label{fig:ComOh}
\end{figure}

\subsection{Communication overhead}\label{sec:sim:comoh}
For each simulated scenario, we now set the CPU resources to 10\% above the amount of resources required by LBound to find a feasible solution when 100\% of the requests are RT. While maintaining this amount of CPU resources, we vary the ratio of RT requests, and measure the overall amount of data for signaling used by \alg. 
Fig.~\ref{fig:ComOh}  presents the per-request signaling overhead, defined as the overall signaling data 
exchanged by \alg\ (``control bytes") over the overall number of critical/new requests along the trace. We consider several values of the \bu\ accumulation delay parameter $\accDelayBu$. A run of \pd() may migrate also non-critical MSs, thus incurring a higher overhead than \pu(). Hence, we set $\accDelayPd = 4 \cdot \accDelayBu$. 
The results show that increasing the fraction of RT requests decreases the signaling overhead. The reason is that RT requests can be placed only in the three lowest levels in the tree, thus avoiding any signaling message between higher-level datacenters. When increasing the accumulation delays, the protocol aggregates more requests before sending a message, thus decreasing the signaling overhead. 
However, an accumulation delay of about $10~\mu\text{s}$ suffices. 
We stress that the accumulation delay only impacts the time until the protocol finds a new feasible placement, not the delay experienced by the user's application. Indeed, delaying the migration decision may deteriorate the user's QoE. However, this performance deterioration can be mitigated using an efficient prediction mechanism for the user's mobility. Furthermore, in practical scenarios, an accumulation delay of about $10~\mu\text{s}$ may be negligible compared to the more considerable delay incurred by the migration process.  
Finally, in all the settings considered, the signaling overhead associated with each request is about $100$~bytes. Thus, we can claim a very low overhead for \alg.

\subsection{Cost comparison}\label{sec:sim:cost}

In our next experiment, we compare the cost of various solutions for the \migProb. We set the ratio of RT requests to 30\%, and vary the resource augmentation. 

The results (not detailed here due to lack of space) show that the costs obtained by BUPU and \alg\ are almost identical, and both are up to 10\% higher than LBound (depending on the concrete setting). That is, despite being distributed and asynchronous, \alg\ obtain costs that are almost identical to those obtained by BUPU, which relies on a centralized controller with an always-accurate view of the system state. \ffit\ typically only finds {\em any} feasible solution when resources are abundant, in which case all the placement algorithms easily obtain close-to-optimal costs.

\section{Related Work}\label{sec:related_work}

State-of-the-art solutions to the  \migProb~\cite{Companion_Fog_short, dynamic_sched_and_reconf_t_short, SFC_mig_short, Avatar}.
assume a centralized orchestrator that possesses a fresh, accurate information about the locations, trajectories, and computational demands of all users, and the available resources at all the datacenters. Such an assumption may be impractical in a large system, possibly operated by several distinct operators.

Other solutions~\cite{Companion_Fog_short, NFV_ego_learning} independently select an optimal destination for each migration request, based on multiple considerations, such as topological distance, availability of resources at the destination, and the data protection level in the destination. 
However, such a selfish user-centric approach may fail to provide a feasible system-level solution when multiple RT users compete for resources in the edge.  

The work~\cite{App_placement_in_fog_n_edge} uses dynamic clustering of datacenters to handle multiple simultaneous independent placement requests. However, the complex dynamic clustering mechanism may result in significant communication and computational overhead. Also,~\cite{App_placement_in_fog_n_edge} does not consider migrating non-critical requests to make room for a new user, as we do. 

The \migProb\ combines several properties of the Multiple Knapsack problem~\cite{MultiKnap} with added restrictions typical for bin-packing problems (e.g., each item can be packed only on a subset of the knapsacks, and a feasible solution must pack all the items). However, in contrast to the usual settings of such problems, we aim at a distributed and asynchronous scheme that runs independently on multiple datacenters (``knapsacks''), using only little communication between them.

The work~\cite{Dance_elephants_VM_mig_short} optimizes lower-level implementational details of the migration process, to decrease its overhead. 
LSTM~\cite{LSTM_predict_and_then_optimize_short} considers learning algorithms that predict future service requests. These solutions are orthogonal to the \migProb\ and hence could be incorporated into our solution to boost performance.

\section{Conclusions}\label{sec:conclusion}
We proposed a distributed asynchronous protocol for service provisioning in the cloud-edge continuum. 
Our solution is carefully designed to reduce signaling overhead by using only small control messages between immediate neighbor datacenters. Numerical results, derived using realistic settings, show that our approach may provide a feasible solution while using only slightly higher computing resources than a centralized scheduler, which may be impractical for large communication networks. Also, our protocol obtains reduced costs and incurs only a small communication overhead.

\bibliographystyle{IEEEtran}
\bibliography{Refs}

\begin{thebibliography}{10}
\providecommand{\url}[1]{#1}
\csname url@samestyle\endcsname
\providecommand{\newblock}{\relax}
\providecommand{\bibinfo}[2]{#2}
\providecommand{\BIBentrySTDinterwordspacing}{\spaceskip=0pt\relax}
\providecommand{\BIBentryALTinterwordstretchfactor}{4}
\providecommand{\BIBentryALTinterwordspacing}{\spaceskip=\fontdimen2\font plus
\BIBentryALTinterwordstretchfactor\fontdimen3\font minus
  \fontdimen4\font\relax}
\providecommand{\BIBforeignlanguage}[2]{{%
\expandafter\ifx\csname l@#1\endcsname\relax
\typeout{** WARNING: IEEEtran.bst: No hyphenation pattern has been}%
\typeout{** loaded for the language `#1'. Using the pattern for}%
\typeout{** the default language instead.}%
\else
\language=\csname l@#1\endcsname
\fi
#2}}
\providecommand{\BIBdecl}{\relax}
\BIBdecl

\bibitem{Justifies_path_to_root_n_CLP_vehs_short}
B.~Kar \emph{et~al.}, ``{QoS} violation probability minimization in federating
  vehicular-fogs with cloud and edge systems,'' \emph{IEEE Transactions on
  Vehicular Technology}, vol.~70, no.~12, pp. 13\,270--13\,280, 2021.

\bibitem{SFC_mig_short}
D.~Zhao \emph{et~al.}, ``Mobile-aware service function chain migration in
  cloud--fog computing,'' \emph{Future Generation Computer Systems}, vol.~96,
  pp. 591--604, 2019.

\bibitem{orch_cloud2edge_survey_short}
S.~Svorobej \emph{et~al.}, ``Orchestration from the cloud to the edge,''
  \emph{The Cloud-to-Thing Continuum}, pp. 61--77, 2020.

\bibitem{Justify_CLP_tree_ISPs}
Y.-D. Lin, C.-C. Wang, C.-Y. Huang, and Y.-C. Lai, ``Hierarchical cord for
  {NFV} datacenters: resource allocation with cost-latency tradeoff,''
  \emph{IEEE Network}, vol.~32, no.~5, pp. 124--130, 2018.

\bibitem{tong2016_justify_path_to_root}
L.~Tong, Y.~Li, and W.~Gao, ``A hierarchical edge cloud architecture for mobile
  computing,'' in \emph{IEEE INFOCOM}, 2016, pp. 1--9.

\bibitem{Dynamic_Service_Provisioning_ToN_short}
I.~Cohen \emph{et~al.}, ``Dynamic service provisioning in the edge-cloud
  continuum with bounded resources,'' \emph{IEEE Transaction on Networking}, in
  press, 2023.

\bibitem{SFC_mig_mechanism}
H.~Yu, J.~Yang, and C.~Fung, ``Elastic network service chain with fine-grained
  vertical scaling,'' in \emph{IEEE GLOBECOM}, 2018, pp. 1--7.

\bibitem{dynamic_sched_and_reconf_t_short}
I.~Leyva-Pupo \emph{et~al.}, ``Dynamic scheduling and optimal reconfiguration
  of {UPF} placement in {5G} networks,'' in \emph{ACM MSWiM}, 2020, pp.
  103--111.

\bibitem{Avatar}
X.~Sun and N.~Ansari, ``{PRIMAL}: Profit maximization avatar placement for
  mobile edge computing,'' in \emph{IEEE ICC}, 2016, pp. 1--6.

\bibitem{APSR_short}
I.~Cohen \emph{et~al.}, ``Parallel {VM} deployment with provable guarantees,''
  in \emph{IFIP Networking}, 2021, pp. 1--9.

\bibitem{Crosshaul}
A.~De~La~Oliva \emph{et~al.}, ``Final 5g-crosshaul system design and economic
  analysis,'' \emph{5G-Crosshaul public deliverable}, 2017.

\bibitem{NFV_ego_learning}
T.~Ouyang \emph{et~al.}, ``Adaptive user-managed service placement for mobile
  edge computing: An online learning approach,'' in \emph{IEEE INFOCOM}, 2019,
  pp. 1468--1476.

\bibitem{Mig_in_Mobile_Edge_Clouds_short}
S.~Wang \emph{et~al.}, ``Dynamic service migration in mobile edge-clouds,'' in
  \emph{IEEE IFIP Networking}, 2015, pp. 1--9.

\bibitem{OKPI_short}
Mart{\'\i}n-P{\'e}rez \emph{et~al.}, ``{OKpi}: {All-KPI} network slicing
  through efficient resource allocation,'' in \emph{IEEE INFOCOM}, 2020, pp.
  804--813.

\bibitem{Dynamic_user_demands}
M.~Nguyen, M.~Dolati, and M.~Ghaderi, ``Deadline-aware {SFC} orchestration
  under demand uncertainty,'' \emph{IEEE Transactions on Network and Service
  Management}, pp. 2275--2290, 2020.

\bibitem{Luxembourg_short}
L.~Codec\'a \emph{et~al.}, ``Luxembourg {SUMO} traffic ({LuST}) scenario:
  Traffic demand evaluation,'' \emph{IEEE Intelligent Transportation Systems
  Magazine}, pp. 52--63, 2017.

\bibitem{Monaco}
L.~Codeca and J.~H{\"a}rri, ``Monaco {SUMO} traffic ({MoST}) scenario: A {3D}
  mobility scenario for cooperative {ITS},'' \emph{EPiC Series in Engineering},
  vol.~2, pp. 43--55, 2018.

\bibitem{OpenCellid}
``Opencellid,'' https://opencellid.org/, accessed on 3.10.2021.

\bibitem{App_placement_in_fog_n_edge}
M.~Goudarzi, M.~Palaniswami, and R.~Buyya, ``A distributed application
  placement and migration management techniques for edge and fog computing
  environments,'' in \emph{IEEE FedCSIS}, 2021, pp. 37--56.

\bibitem{SUMO2018_short}
P.~Alvarez \emph{et~al.}, ``Microscopic traffic simulation using sumo,'' in
  \emph{IEEE International Conference on Intelligent Transportation Systems},
  2018.

\bibitem{SFC_mig_Github}
\BIBentryALTinterwordspacing
``Service function chains migration.'' [Online]. Available:
  \url{https://github.com/ofanan/SFC\_migration}
\BIBentrySTDinterwordspacing

\bibitem{omnet}
\BIBentryALTinterwordspacing
``{OMNeT++} discrete event simulator,'' 2023. [Online]. Available:
  \url{https://omnetpp.org}
\BIBentrySTDinterwordspacing

\bibitem{dist_SFC_mig_Github}
\BIBentryALTinterwordspacing
``Distributed {SFC} migration.'' [Online]. Available:
  \url{https://github.com/ofanan/Distributed\_SFC\_migration}
\BIBentrySTDinterwordspacing

\bibitem{Gurobi}
\BIBentryALTinterwordspacing
``Gurobi optimizer reference manual,'' 2023. [Online]. Available:
  \url{https://www.gurobi.com}
\BIBentrySTDinterwordspacing

\bibitem{Companion_Fog_short}
C.~Puliafito \emph{et~al.}, ``Companion fog computing: Supporting things
  mobility through container migration at the edge,'' in \emph{IEEE SMARTCOMP},
  2018, pp. 97--105.

\bibitem{MultiKnap}
M.~S. Hung and J.~C. Fisk, ``An algorithm for 0-1 multiple-knapsack problems,''
  \emph{Naval Research Logistics Quarterly}, vol.~25, no.~3, pp. 571--579,
  1978.

\bibitem{Dance_elephants_VM_mig_short}
K.~Ha \emph{et~al.}, ``You can teach elephants to dance: Agile {VM} handoff for
  edge computing,'' in \emph{ACM/IEEE SEC}, 2017, pp. 1--14.

\bibitem{LSTM_predict_and_then_optimize_short}
T.~Subramanya and R.~Riggio, ``Centralized and federated learning for
  predictive {VNF} autoscaling in multi-domain {5G} networks and beyond,''
  \emph{IEEE TNSM}, vol.~18, no.~1, pp. 63--78, 2021.

\end{thebibliography}

\end{document}